\documentclass[aps,amsmath,twocolumn,amssymb,titlepage,10pt]{revtex4-1}
\usepackage[T1]{fontenc}
\usepackage[utf8]{inputenc}
\usepackage{amsmath}
\usepackage{braket}
\usepackage{amsfonts}
\usepackage{comment}
\usepackage{graphicx}
\usepackage{hyperref}
\usepackage[capitalize]{cleveref}
\usepackage{amssymb}
\usepackage{amsmath}
\usepackage{mathtools}
\usepackage{todonotes}

\begin{document}
\title{Thermodynamics of the metal-insulator transition in the extended\\ Hubbard model from determinantal quantum Monte Carlo}
\author{Alexander Sushchyev}
\author{Stefan Wessel}
\affiliation{Institute for Theoretical Solid State Physics, RWTH Aachen University, JARA Fundamentals of Future Information Technology, and \\ JARA Center for Simulation and Data Science, 52056 Aachen, Germany}
\begin{abstract}
Using finite-temperature determinantal quantum Monte Carlo simulations, we examine the thermodynamic
properties of the extended Hubbard model on the half-filled square lattice in the Slater regime at intermediate coupling. We consider both the case of nearest-neighbor interactions and  long-range Coulomb interactions, for coupling strengths in which the presence of  non-local interactions  still allows us to perform  sign-problem free quantum Monte Carlo simulations. In particular, we assess a recently proposed scenario from variational calculations in terms of  a first-order metal-insulator transition in this interaction regime.
\end{abstract}
\maketitle
\section{Introduction}\label{Sec:Introduction}
The Hubbard model~\cite{Hubbard63}, describing itinerant electrons in the presence of a local (onsite) repulsion, provides a most basic model to study the competition between kinetic energy and interaction effects in fermionic quantum many-body systems. A wide range of theoretical and computational approaches have been employed in order to explore its physical properties and its relevance to a wide breath of fundamental phenomena in condensed matter physics has been demonstrated, including the Mott-insulator transition, and the emergence of symmetry broken states, such as antiferromagnetism (AFM) or superconductivity (see, e.g., Ref.~\cite{Arovas22} for a recent review).  This effort has contributed substantially to our current understanding of strongly correlated electron systems. Moreover, cold atom experiments provide us with a unique platform to study the physics of the Hubbard model over a wide range of controllable parameters~\cite{Esslinger10}. 

However, in view of the fact that in solid-state materials the long-range Coulomb (LRC) interaction is only partially screened, it is  important to also account for the effects of more extended interactions in addition to the local Hubbard-$U$ on the physical properties. Indeed, non-local interactions  affect various quantities such as the electronic band width~\cite{Ayral17,InTVeld19}, and they can induce charge density wave states~\cite{Bari71,Zhang89,Terletska17,Terletska18,Paki19}, to name but  a few consequences.

Recently, the effects of non-local interactions  on the metal-insulator transition on the half-filled square lattice have been explored, based on a variational approach~\cite{Schueler18, Schueler19}. 
More specifically, by means of  the Peierls-Feynmann-Bogoliubov variational principle, the extended Hubbard model was  approximated in Ref.~\cite{Schueler19} by an effective  (local) Hubbard model in terms of an effective hopping parameter and local interaction strength $\tilde{U}$. For this purpose, the variational free energy was calculated based on the  integration of thermodynamic data for the effective (local) Hubbard model, obtained using determinantal quantum Monte Carlo (DQMC) simulations on a dense parameter grid, applying a two-dimensional Savitzky-Golay filter and spline interpolation to the grid data~\cite{Schueler19}.  
Based on this variational approach, several conclusions regarding the effects of non-local interactions  were drawn. In particular, two distinct mechanisms are described in Ref.~\cite{Schueler19}, how non-local interactions suppress correlation effects: Within the Fermi-liquid regime they  reduce   $\tilde{U}$, while they increase the effective hopping strength within the insulating regime. Moreover,
the competition between both mechanisms was found to drive a first-order  metal-insulator transition in the presence of non-local interactions. From a comparison of the associated entropy jump across the transition with available experimental data on materials with  purely electronic metal-insulator transitions, the authors conclude that non-local interactions are at least in part responsible for the discontinuous metal-insulator transitions observed in  correlated electron materials. 
In view of the above, it is mandatory to compare the results from the variational approach to unbiased calculations that take the non-local interactions fully into account. In fact, as  noted in Ref.~\cite{Schueler19}, the  parameter region in which the discontinuous thermodynamic behavior was observed is accessible to sign-problem free DQMC simulations, i.e., including a full treatment of the non-local interaction terms~\cite{Hohenadler14,Golor15}. 

Here, we report results from unbiased DQMC simulations of the extended Hubbard model in order to assess the qualitative and quantitative appropriateness of the variational approach~\cite{Schueler19}. In particular, we probe for unbiased evidence for the emergence of the discontinuous thermodynamic behavior reported in 
Ref.~\cite{Schueler19} as a result of  non-local interactions. 
The remainder of this paper is organized as follows: In Sec.~\ref{Sec:Model} we define the extended Hubbard models that we consider here, and  also specify our  computational approach.  The results obtained from our DQMC calculations are then reported in Sec.~\ref{Sec:Results}, and final conclusions are given in Sec.~\ref{Sec:Conclusions}.

\section{Model and Method}\label{Sec:Model}
In the following, we consider the extended Hubbard model with non-local density-density interactions, described by the Hamiltonian
\begin{equation}
    H=-t\!\!\sum_{\langle i,j \rangle, \sigma} \!\! (c^\dagger_{i\sigma} c_{j\sigma}+ \mathrm{h.c.} )
      +U\sum_i n_{i\uparrow}n_{j\downarrow}
      +\frac{1}{2}\sum_{i\neq j} V_{ij} n_i n_j, 
\end{equation}
on the square lattice. Here, $c^\dagger_{i\sigma}$ ($c_{i\sigma}$) denotes the creation (annihilation) operator for electrons on site $i$ with spin projection $\sigma$,  $n_{i\sigma}=c^\dagger_{i\sigma} c_{i\sigma}$ the local spin-resolved occupation operators, and 
$n_i=n_{i\uparrow}+n_{i\downarrow}$ the total local occupation. Furthermore, $t$ denotes the  nearest-neighbor hopping amplitude, $U$ the local (Hubbard) repulsion, and $V_{ij}$ the non-local interaction between electrons on site $i$ and $j$. In the following, we will focus on two different cases: (i) the so-called $U$-$V$ model with nearest neighbor interactions of strength $V$, and (ii) LRC interactions with $V_{ij}=V_C/ d_{ij}$, where $d_{ij}$ denotes the distance between sites $i$ and $j$ 
(the lattice constant $a=1$).
In all our investigations, we consider the case of  half-filling.

In order to examine the thermodynamic properties of the above model, we use DQMC simulations~\cite{Blankenbecler81}, performed using the ALF code~\cite{Bercx13}. This allows us to simulate the above model sign-problem free using appropriate Hubbard-Stratanovich 
decoupling schemes~\cite{Hohenadler14,Golor15}  for both the onsite and non-local interactions within the regimes (i) $V\leq  U/4$ for the $U$-$V$ model, and (ii) $V_C \lesssim 0.62 U$, such that $U\delta_{ij}+V_{ij}$ is a positive-definite matrix, for the LRC-Hubbard model case. Note that an earlier DQMC work on the $U$-$V$ extended Hubbard model used a decoupling scheme that leads to a sign-problem for any finite $V>0$~\cite{Zhang89} (this fact is not mentioned explicitly in that reference). For the DQMC simulations, we consider finite square lattices with $N=L\times L$ lattice sites with periodic boundary conditions in both lattice directions and the standard minimum image convention for the case of LRC interactions. We denote the (inverse) temperature by $T$ ($\beta=1/T$) in the following ($k_B=1$). 

In our analysis, we mainly  concentrate on  DQMC results for the double occupancy, 
\begin{equation}
    D=\frac{1}{N} \sum_{i=1}^N \langle n_{i\uparrow} n_{i\downarrow} \rangle,
\end{equation}
for which we perform a spatial averaging  in order to enhance the statistical accuracy. We  also draw attention to a recent proposal on how to improve the DQMC sampling procedure in order to further reduce statistical fluctuations on such local quantities~\cite{Ulybyshev21}.
For the DQMC simulations a  Trotter-decomposition of $H$ is used with a small imaginary-time step $\Delta\tau$. For the reported DQMC results for $D$, we preformed a $\Delta\tau\rightarrow 0$ extrapolation, as detailed in App.~\ref{App:A}, and the data shown below has  always been obtained from this analysis.
Furthermore, we  report results for the entropy $S$ (per site), which we obtain from the DQMC values of the internal energy $E$ (per site) at $\Delta\tau=0.1/t$  via  thermodynamic integration, 
\begin{equation}
    S(\beta)= \beta E(\beta) + \ln(4) - \int_0^{\beta} E(\beta')\, d\beta'.
\end{equation}
The integral is evaluated numerically using the  trapezoidal rule on a dense $\beta$-mesh. In addition to these thermodynamic quantities, we also measured in the QMC simulations the structure factors for 
AFM, stabilized, e.g.,  in the ground state of the Hubbard model at half-filling, 
\begin{equation}
    S_\text{AF}=\frac{1}{N}\sum_{i,j=1}^N \varepsilon_i \varepsilon_j \langle \mathbf{S}_i\cdot \mathbf{S}_j\rangle,
\end{equation}
as well as for the commensurate charge density wave (CDW) state that is expected to be stabilized for sufficiently  strong $V$ in the $U$-$V$ model~\cite{Zhang89},  
\begin{equation}
    S_\text{CDW}=\frac{1}{N}\sum_{i,j=1}^N \varepsilon_i \varepsilon_j \langle n_i n_j\rangle.
\end{equation}
Here, $\varepsilon_i=\pm 1$, depending on which sublattice the site $i$ belongs to on the bipartite square lattice.

\section{Results}\label{Sec:Results}
In this section, we report  results from DQMC simulations and compare them with previously reported findings. 
Before we consider the case of non-local interactions, we first present DQMC results for the Hubbard model (i.e., $V=V_C=0$).

\subsection{The Hubbard model}\label{Sec:Hubbard}

\begin{figure}[t]
    \centering
    \includegraphics{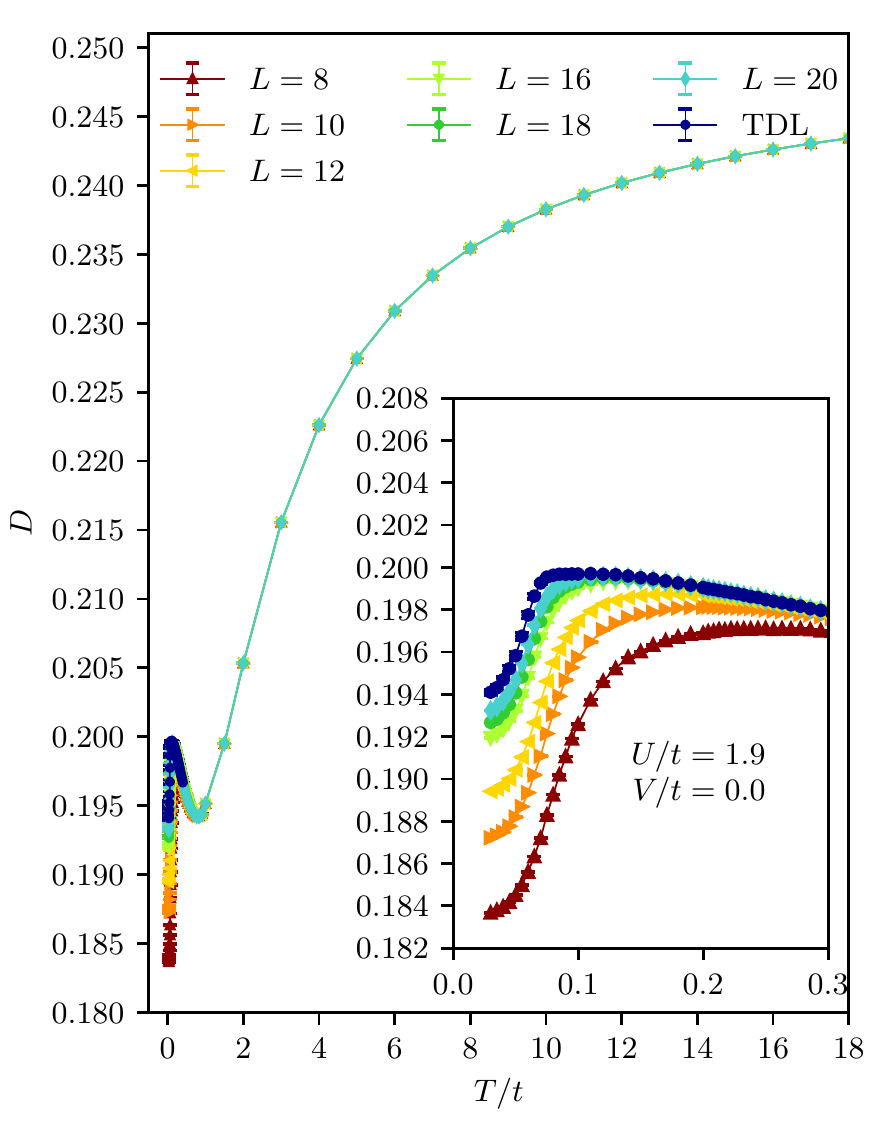}
    \caption{Temperature dependence of the double occupancy $D$ for the Hubbard model at $U/t=1.9$. The inset focuses on the low-temperature regime containing the local maximum.}
    \label{Fig:Hubbard_D}
\end{figure}

\begin{figure}[t]
    \centering
    \includegraphics{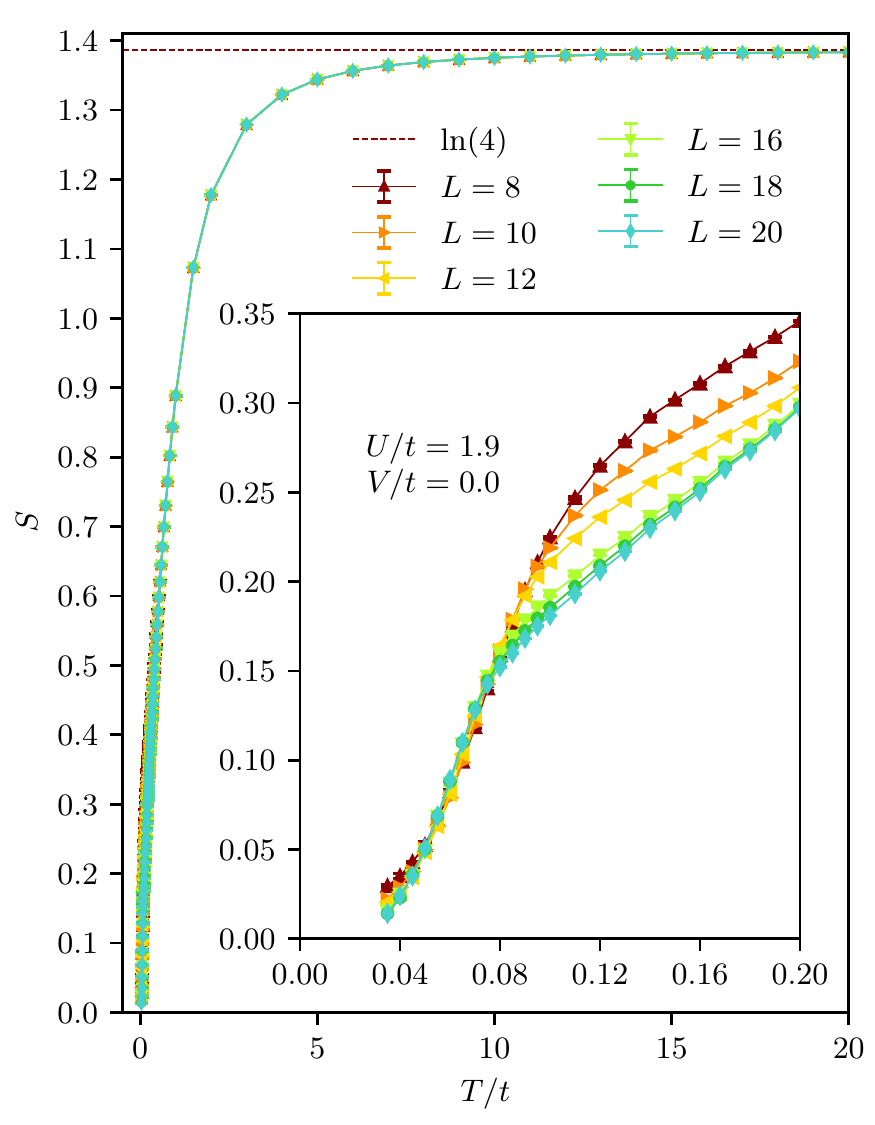}
    \caption{Temperature dependence of the entropy $S$ for the Hubbard model at $U/t=1.9$. The inset focuses on the low-temperature regime.}
    \label{Fig:Hubbard_S}
\end{figure}

While highly accurate and detailed DQMC results  of various thermodynamic quantities for the Hubbard model has been reported in Ref.~\cite{Fratino17}, these do not include the thermal behavior of both $D$ and $S$, which we  report below.
In order to allow for a direct comparison to the numerical results reported from the variational approach in Ref.~\cite{Schueler19}, we focus in the following on the specific parameter ratio $U/t=1.9$, which locates the electronic system within the Slater regime. A characteristic feature of $D$ in this regime is the appearance of a maximum at low temperatures.  In Fig.~\ref{Fig:Hubbard_D}, we show the DQMC results for $D$ as a function of temperature as obtained for various system sizes $L$. Also included is the result from an extrapolation to the thermodynamic limit (TDL), performed as detailed in App.~\ref{App:B}. 

A prominent feature in the Slater regime is the non-monotonous behavior of $D$ in Fig.~\ref{Fig:Hubbard_D},  displaying both a local minimum and maximum. We quickly review, how this behavior comes about. A   minimum in $D$ is  observed also in cases where the long-range AFM order in the ground state is quenched (e.g., by geometric frustration~\cite{Laubach15} or within the non-magnetic dynamical-mean-field-theory  approximation of the Hubbard model~\cite{Georges92}). It  results from the initial decrease of $D$ with increasing $T$ via an entropic effect, 
in  analog of the Pomeranchuk effect in  Helium 3 or ultracold atoms~\cite{Werner05}. Namely, the formation of
non-ordered local magnetic moments leads to an entropy gain with respect to a state of itinerant electrons. Upon further increasing $T$ however, $D$  eventually needs to increase again towards the non-interacting value of $1/4$ at infinite temperature. Taken together, this results in the local minimum of $D$ that is visible in 
Fig.~\ref{Fig:Hubbard_D}, with  $T_\mathrm{min}=0.83 t$ obtained from the DQMC simulations.  

The formation of a local maximum in $D$ in Fig.~\ref{Fig:Hubbard_D} at $T_\mathrm{max}= 0.085 t$ results from the proliferation of AFM fluctuations for the Hubbard model on the bipartite square lattice via an energetic effect: In the Slater regime,  the onset of AFM fluctuations lead to a decrease of the local Hubbard interaction energy $UD$ when decreasing $D$ upon further lowering the temperature~\cite{Fratino17}. 
Two mechanism are thus responsible for the low-temperature maximum appearing in $D$: Moving away  from the maximum towards lower temperature, $D$ is lowered to decrease the interaction energy (Slater effect), while $D$ is suppressed towards higher temperature to increase the spin entropy (Pomeranchuk effect). We note that the DQMC data in Fig.~\ref{Fig:Hubbard_D} exhibits a more rapid drop of $D$ upon decreasing $T$ from the position of the local maximum at $T_\mathrm{max}$ than upon increasing $T$.

We can directly compare the DQMC data to the results reported from the variational approach~\cite{Schueler19}, where we focus in particular on the local maximum in the low-temperature region. With respect to the position of the maximum, both approaches agree rather well. However, both the finite-size DQMC data and the TDL values of $D$ at the local maximum fall below 0.2, whereas within the variational approach the local maximum extends beyond 0.207. While this difference might be considered small, it is   are actually relevant in view of the fact that the jumps in $D$ reported in Ref.~\cite{Schueler19} for finite $V,V_C>0$ are of even smaller magnitude, as detailed below. 

Before we turn to non-zero values of $V$, we briefly report the DQMC results for the entropy $S$ for the same value of $U/t=1.9$, cf. Fig.~\ref{Fig:Hubbard_S}. In the intermediate temperature regime below about $T\approx t$, we 
observe an essentially linear decrease of $S$ for large systems, down to a temperature of $T\approx 0.08 t$. At lower temperatures, the entropy decreases more rapidly with decreasing temperature. Overall, this behavior matches the aforementioned similar asymmetric drop in $D$ near its local maximum at $T_\mathrm{max}$. The enhanced reduction of both $D$ and $S$ below $T_\mathrm{max}$ reflects the Stoner effect mentioned above upon entering the regime where AFM fluctuations proliferate. In the finite-size data of the entropy we furthermore identify a small temperature window below $T_{\mathrm{max}}$, in which $S$ exhibits a (mild) increase with system size $L$, in contrast to its decrease with increasing $L$ outside  of this regime. A possible explanation for the anomalous behavior is the growth of the AFM correlation length on scales comparable to the simulated system sizes in this regime. By contrast, for lower (higher) temperatures, the correlation length instead resides well beyond (below) the finite-size of the simulation cell.

\subsection{The $U$-$V$ model}\label{Sec:UV}

\begin{figure}[t]
    \centering
    \includegraphics{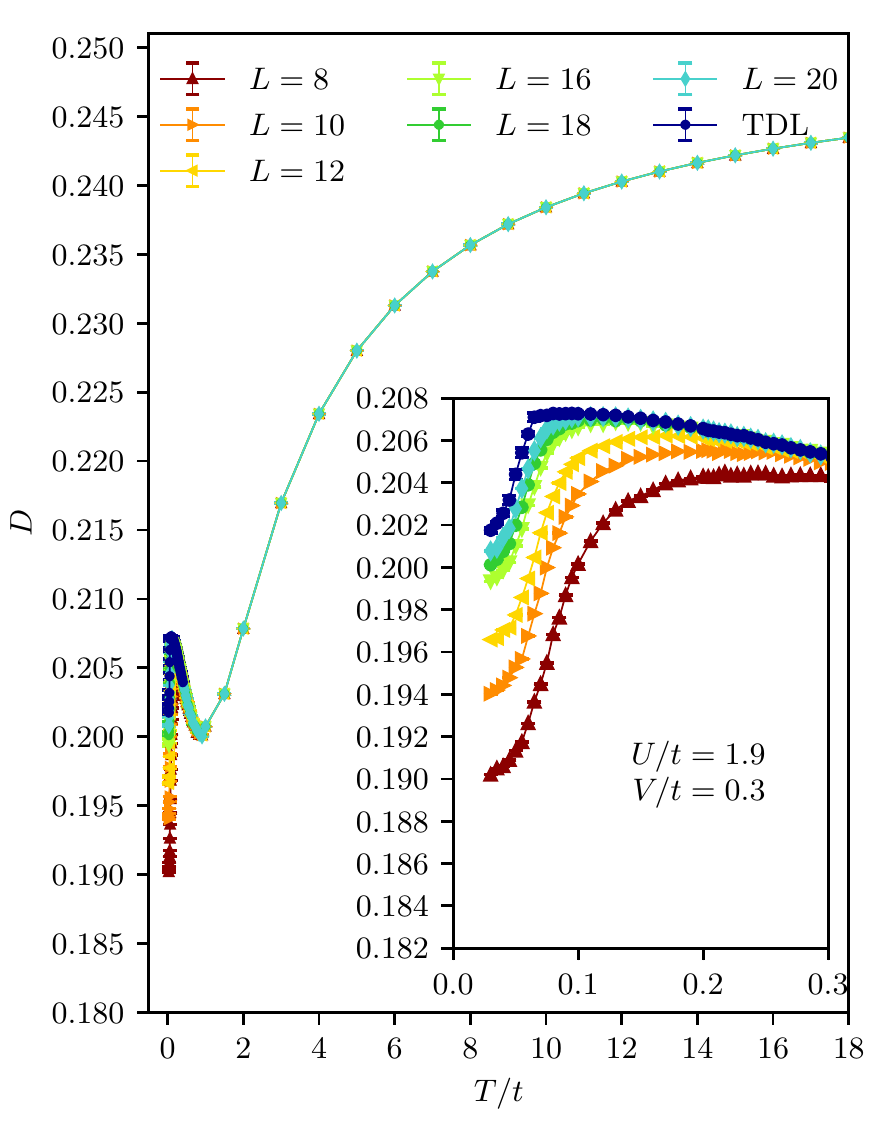}
    \caption{Temperature dependence of the double occupancy $D$ for the $U$-$V$ model at $U/t=1.9$, $V/t=0.3$. The inset focuses on the low-temperature regime containing the local maximum.}
    \label{Fig:UV_D}
\end{figure}

\begin{figure}[t]
    \centering
    \includegraphics{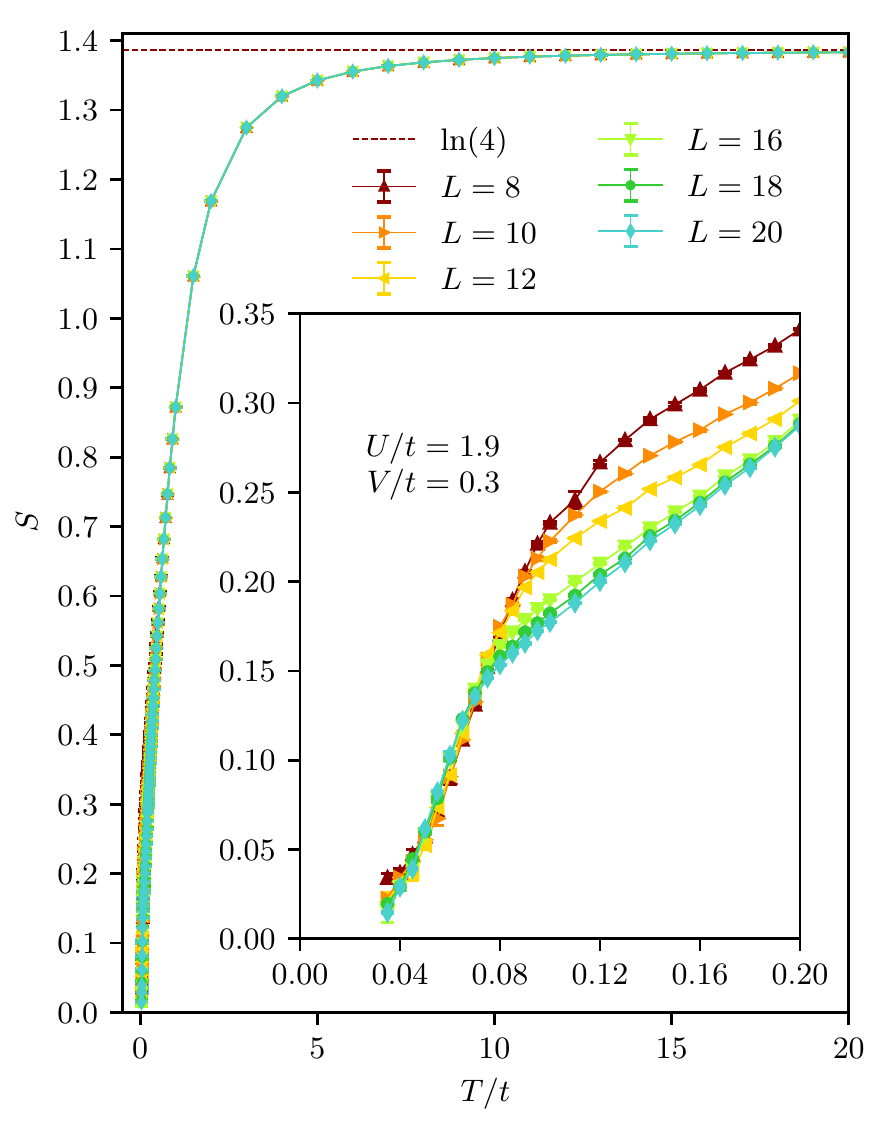}
    \caption{Temperature dependence of the entropy $S$ for the $U$-$V$ model at $U/t=1.9$, $V/t=0.3$. The inset focuses on the low-temperature regime.}
    \label{Fig:UV_S}
\end{figure}

\begin{figure}
    \centering
    \includegraphics{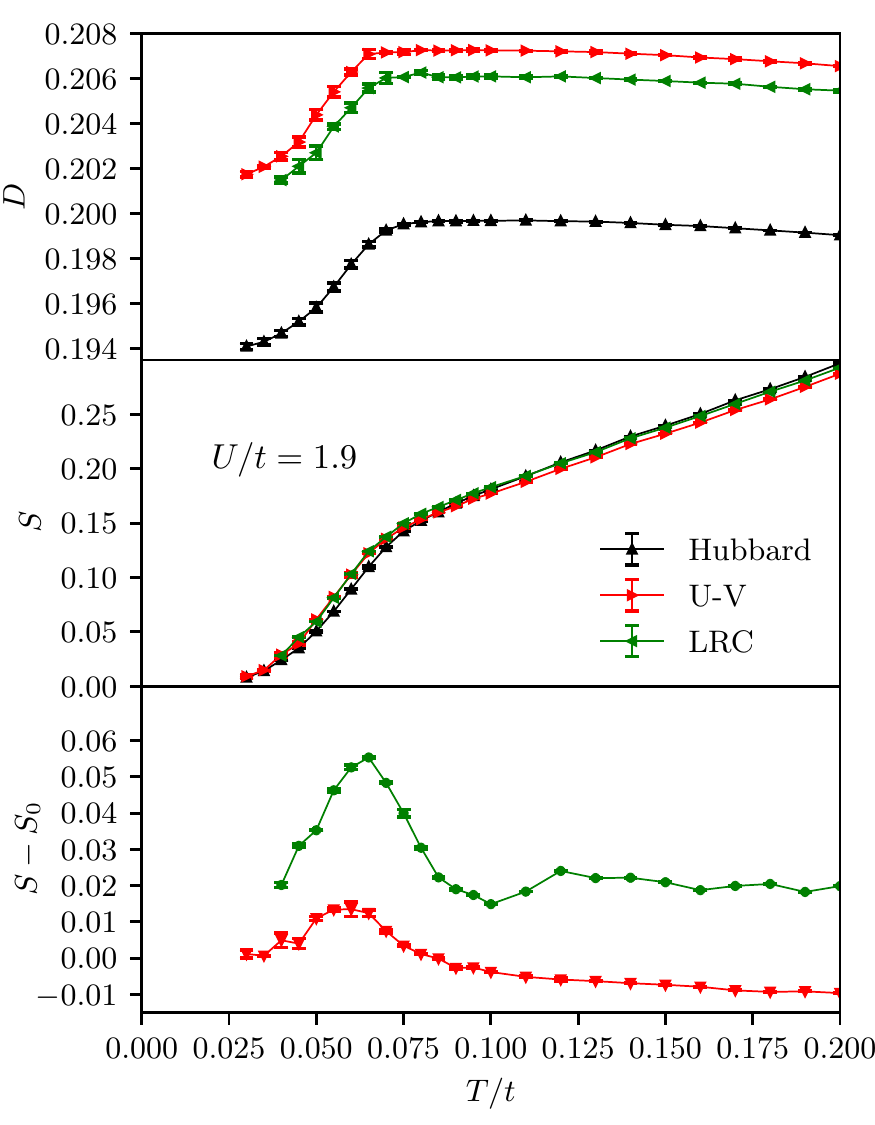}
    \caption{Temperature dependence of the TDL-extrapolated double occupancy $D$ (top panel) and the entropy $S$ for the $L=20$  systems (middle panel) for the considered models with $U/t=1.9$, for  $V/t=0.3$, and $V_C/t=1.9$, respectively. The bottom panel shows the difference $S-S_0$ of the entropy for the models with extended interactions with respect to the entropy of the Hubbard model (denoted $S_0$ here).}
    \label{Fig:comparison}
\end{figure}

We next turn to the $U$-$V$ model, and consider in particular a value of $V/t=0.3$, again for $U/t=1.9$. Namely, for this value of $V$ a noticeable jump in $D$ was observed within the variational approach at a temperature of $T\approx 0.085 t$, and was taken as indication of a first-order metal-insulator transition~\cite{Schueler19}. Performing the data analysis  as in the previous section, we obtain the DQMC results shown in Figs.~\ref{Fig:UV_D} and ~\ref{Fig:UV_S} for $D$ and $S$, respectively. For the purpose of a direct comparison, the QMC data for both $D$ and $S$ for the different models are also collected in Fig.~\ref{Fig:comparison}. In agreement with general expectations and the results from the variational approach, we observe an increase of $D$ for the case of finite $V$ as compared to the $V=0$ case, corresponding to an overall decrease of the local correlations. Correspondingly, we also observe a mild enhancement of the entropy in the low-temperature  region for finite $V>0$, while otherwise $S$ also shows a  behavior similar to the one at  $V=0$. Besides the overall increase in the values of $D$, we observe no significant change in, e.g.,  the temperature of the maximum in $D$.

In contrast to the variational approach,  the finite-size DQMC data does not exhibit any indication for the  onset of a discontinuity in $D$ near the local maximum. We do observe in the TDL extrapolation a somewhat steeper drop of $D$ on the low-temperature side of the maximum than for $V=0$, but no indication for a jump is obtained. 
It was already noted in Ref.~\cite{Schueler19} that the discontinuity obtained within the variational approach is of order $3\times 10^{-4}$ for $V/t=0.3$ and thus actually rather small. 
On the other hand, we noted  already for the case $V=0$, discussed in the previous section, that the values for $D$ obtained in Ref.~\cite{Schueler19} deviate from the DQMC results by much larger differences (of order $7\times 10^{-3})$. Moreover, 
the finite-size systems studied in Ref.~\cite{Schueler19} for $V=0$ extend up to $L=12$, i.e., well below the value of $L$ up to $20$ used here. Together, these observations suggest that the deviations seen between the  DQMC data and the variational approach are due to the interpolation and approximation schemes that were employed in Ref.~\cite{Schueler19}. While we cannot exclude from our analysis, that an extremely weak discontinuity may eventually emerge (e.g.,  for values of $V>U/4$, i.e., outside the regime accessible to DQMC), the above  direct comparison of the magnitude of the deviations obtained from the variational approach with the DQMC  data  for the case of $V=0$ indicates that the weak values of the reported discontinuity actually fall well within the error margins of the variational approach. 

\begin{figure}[t]
    \centering
    \includegraphics{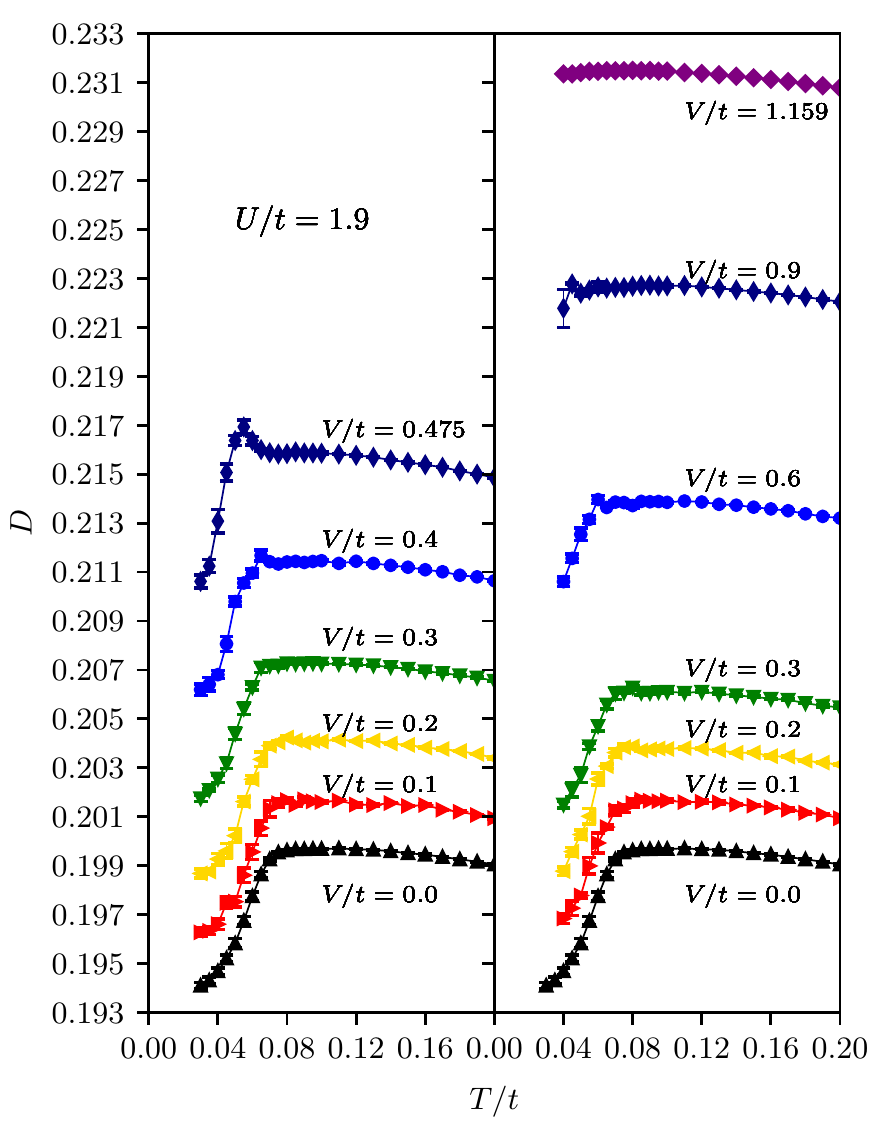}
    \caption{Temperature dependence of the TDL-extrapolated double occupancy $D$ for the $U$-$V$ model (left panel) and the LRC-Hubbard model (right panel) at $U/t=1.9$ for various values of $V$ and $V_C$, respectively.}
    \label{Fig:D_ext}
\end{figure}

\begin{figure}[t]
    \centering
    \includegraphics{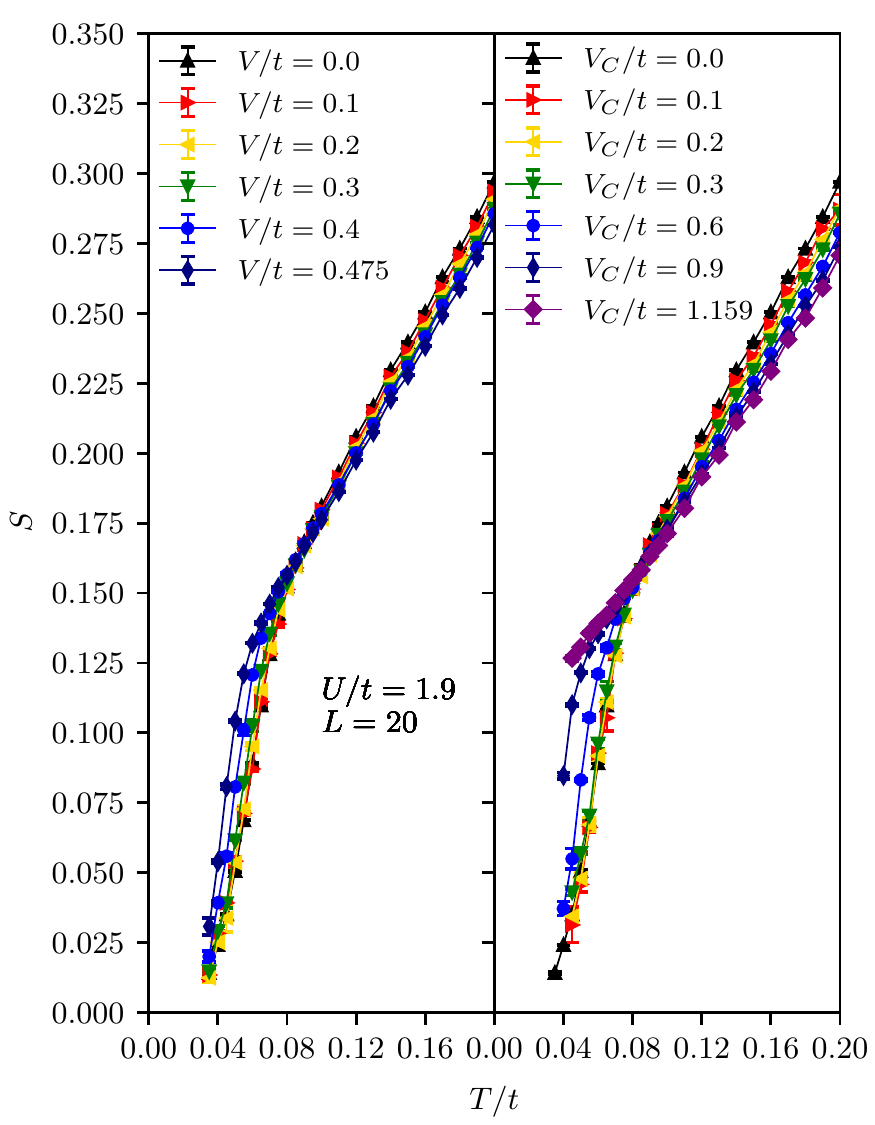}
    \caption{Temperature dependence of the entropy  $S$ for the $U$-$V$ model (left panel) and the LRC-Hubbard model (right panel) at $U/t=1.9$ for various values of $V$ and $V_C$, respectively on the $L=20$ system.}
    \label{Fig:S_ext}
\end{figure}

Thus far, we concentrated on the specific value of $V/t=0.3$.  In addition, we performed DQMC simulations at several other values of $V$, up to and including the limiting case of $V=U/4=0.475t$ for sign-problem free DQMC simulations at $U/t=1.9$. The TDL-extrapolated values of $D$ obtained from these extended  simulations of the $U$-$V$ model are shown in the left panel of Fig.~\ref{Fig:D_ext}, while the left panel of Fig.~\ref{Fig:S_ext} summarizes the results for the entropy $S$ obtained on the largest considered system size ($L=20$) for the $U$-$V$ model.

\begin{figure}[t]
    \centering
    \includegraphics{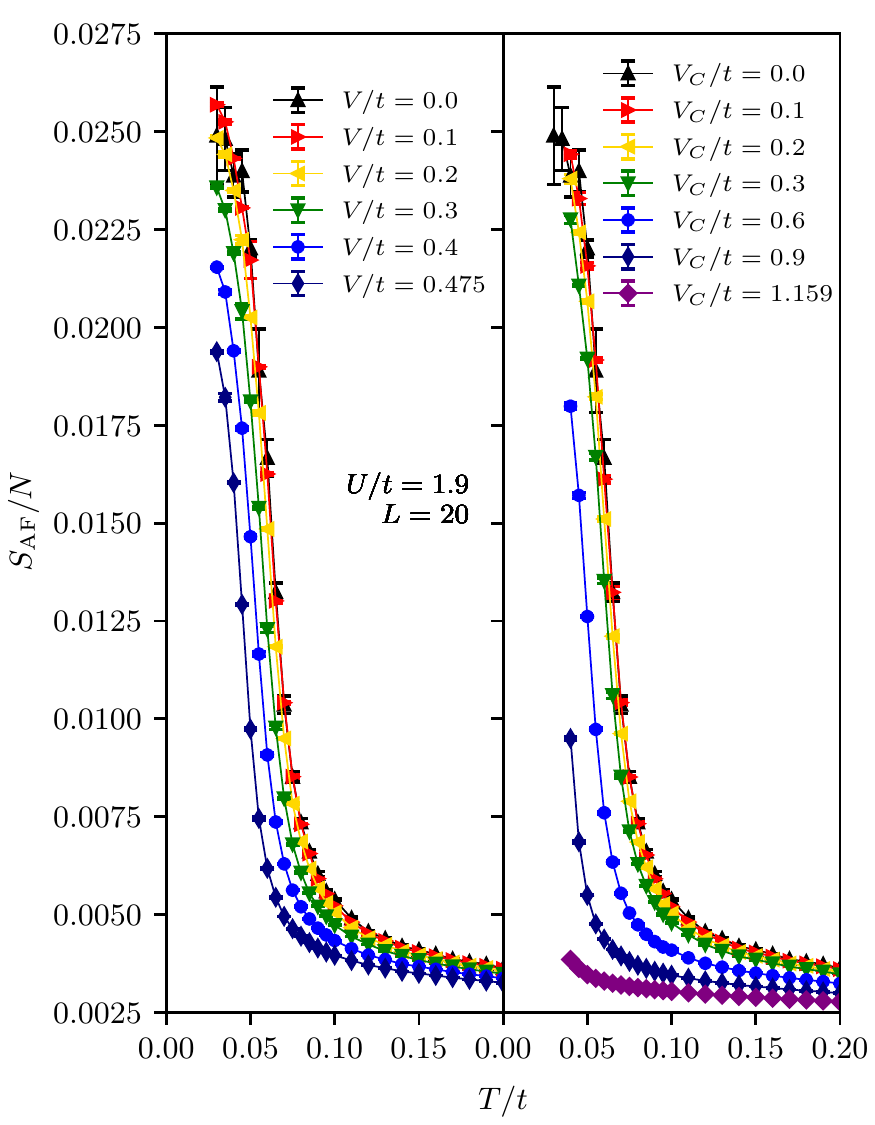}
    \caption{Temperature dependence of the AFM structure factor $S_\mathrm{AF}$ for the $U$-$V$ model (left panel) and the LRC-Hubbard model (right panel) at $U/t=1.9$ for various values of $V$ and $V_C$, respectively on the $L=20$ system.}
    \label{Fig:SAF_ext}
\end{figure}

\begin{figure}[t]
    \centering
    \includegraphics{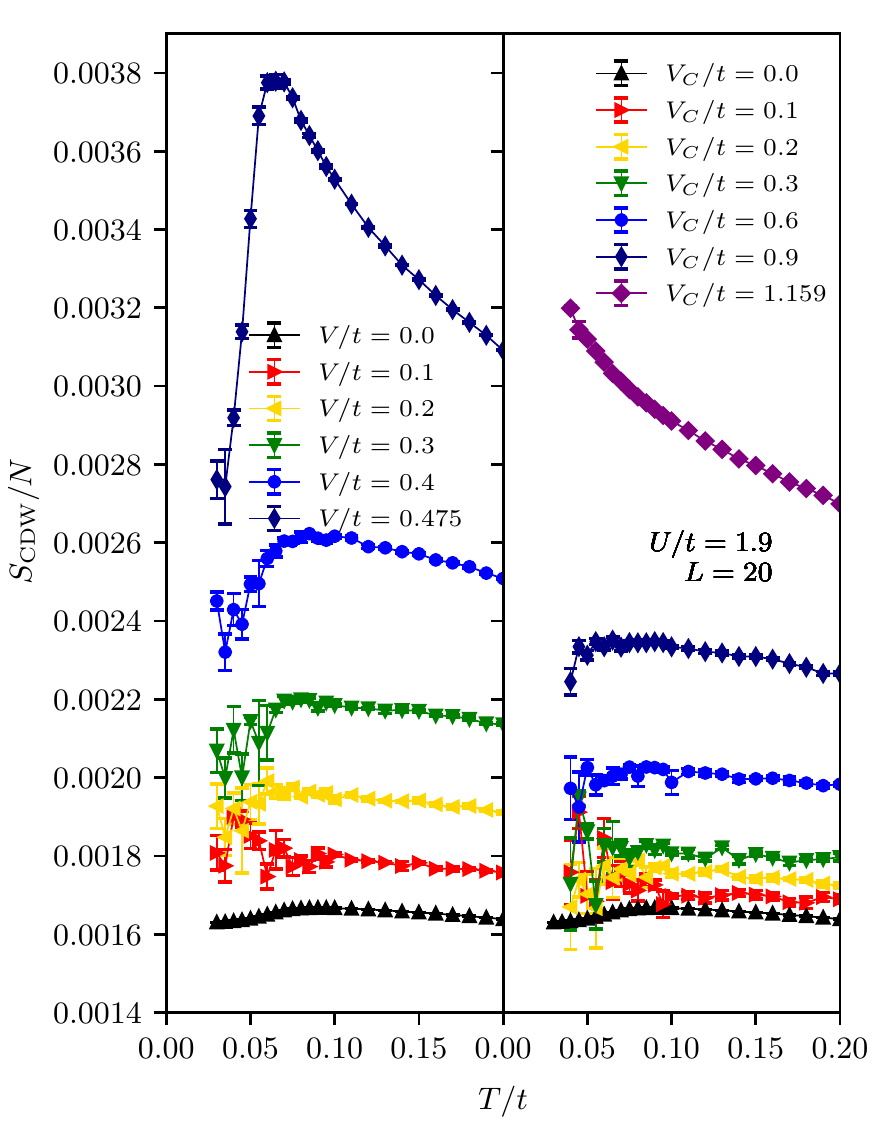}
    \caption{Temperature dependence of the CDW structure factor $S_\mathrm{CDW}$ for the $U$-$V$ model (left panel) and the LRC-Hubbard model (right panel) at $U/t=1.9$ for various values of $V$ and $V_C$, respectively on the $L=20$ system.}
    \label{Fig:SCDW_ext}
\end{figure}

Both quantities exhibit  similar behavior and trends for the various values of $V$ as for the case $V/t=0.3$ considered in detail above --  the most noticeable difference being the presence of a (weak) peak in $D$ at the largest accessible values of $V/t \gtrsim 0.4$. We can understand the corresponding enhancement in the local density fluctuations from considering the ordering tendencies of the $U$-$V$ model: Namely, for sufficiently strong $V$, the $U$-$V$ model is expected to stabilize a CDW ground state~\cite{Zhang89}. Recent zero-temperature DQMC simulations~\cite{Yao22} have shown that for $V\leq U/4$ the system has an AFM ground state, i.e., CDW order sets in for $V$ sufficiently larger than $U/4$ only.
In Fig.~\ref{Fig:SAF_ext} (left panel) and Fig.~\ref{Fig:SCDW_ext} (left panel), we provide DQMC data for the structure factors $S_\mathrm{AF}$ and $S_\mathrm{CDW}$ of the $U$-$V$ model, respectively. Both quantities are shown as functions of temperature on the largest considered system size, $L=20$, and for various values of $V$. We find that within the DQMC-accessible region, the AFM structure factor increases steadily with decreasing $T$, in accord with the AFM ground state in this regime (by the Mermin-Wagner theorem, AFM order is destroyed by thermal fluctuations at any finite temperatures in the TDL). On the other hand, the CDW structure factor, while initially increasing upon lowering the temperature, is eventually suppressed again at low $T$, in accord with the findings in Ref.~\cite{Yao22}, i.e., for $V\leq U/4$ the ground state has AFM but no CDW order. 

In contrast to AFM, the CDW order that emerges for  large $V$ is stable with respect to (weak) thermal fluctuations, i.e., the CDW order stabilized at sufficiently large $V$ melts at a {\it finite} critical temperature across a  thermal phase transition (from symmetry considerations, this transition is continuous and belongs to the two-dimensional Ising universality class). For values of $V$ below but near $U/4$, the critical local density fluctuations associated to this nearby thermal critical region will enhance the double occupancy $D$, an effect that apparently is captured by the data in Fig.~\ref{Fig:D_ext}. From these considerations one may identify the peak position in $D$ at about $T/t\approx 0.06$ as a rough estimate for the critical temperature of the CDW melting transition in the interaction regime where the CDW ground state emerges. 
It would certainly be interesting to explore this regime in more detail by simulating the system for even larger values of $V>U/4$ and to examine the thermal phase diagram in this regime. However, due to the sign problem, this is not feasible within the DQMC approach. 

\subsection{The LRC-Hubbard model }\label{Sec:Coulomb}
\begin{figure}[t]
    \centering
    \includegraphics{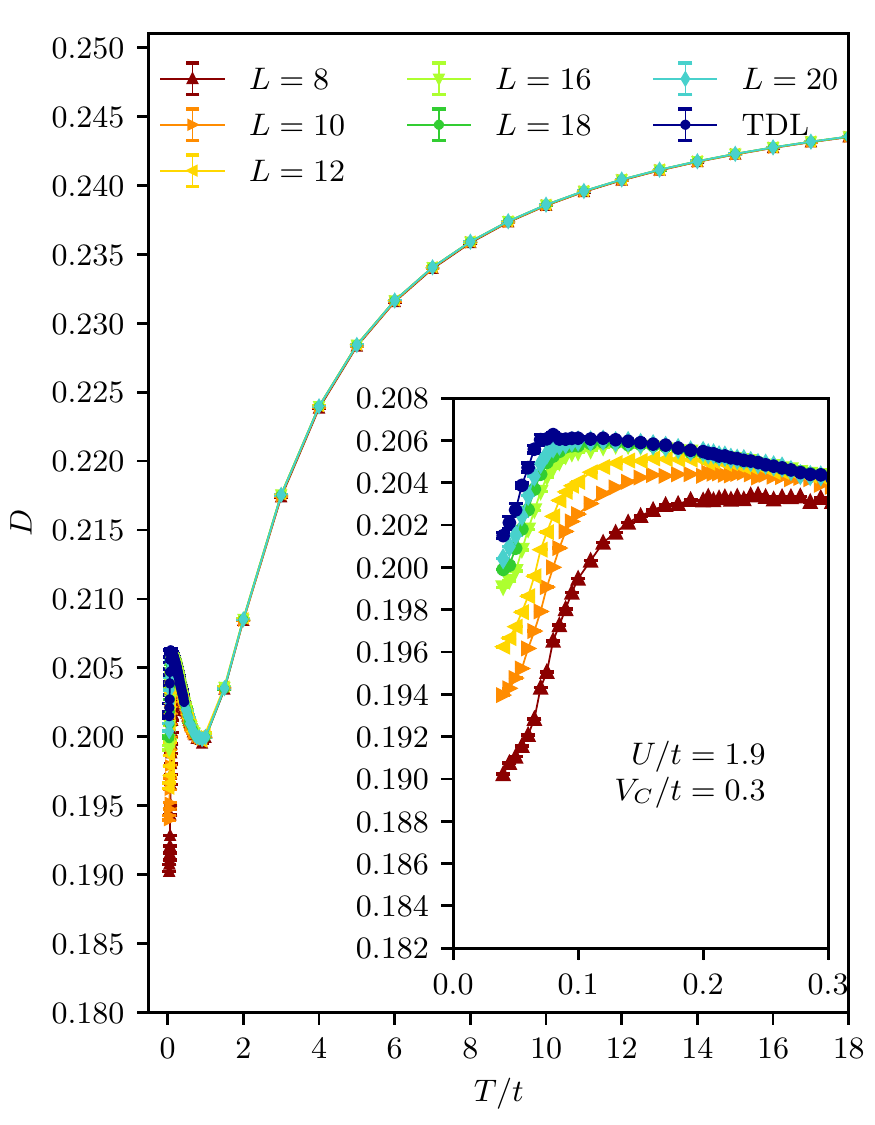}
    \caption{Temperature dependence of the double occupancy $D$ for the LRC-Hubbard model at $U/t=1.9$, $V_C/t=0.3$. The inset focuses on the low-temperature regime containing the local maximum.}
    \label{Fig:Coulomb_D}
\end{figure}

\begin{figure}[t]
    \centering
    \includegraphics{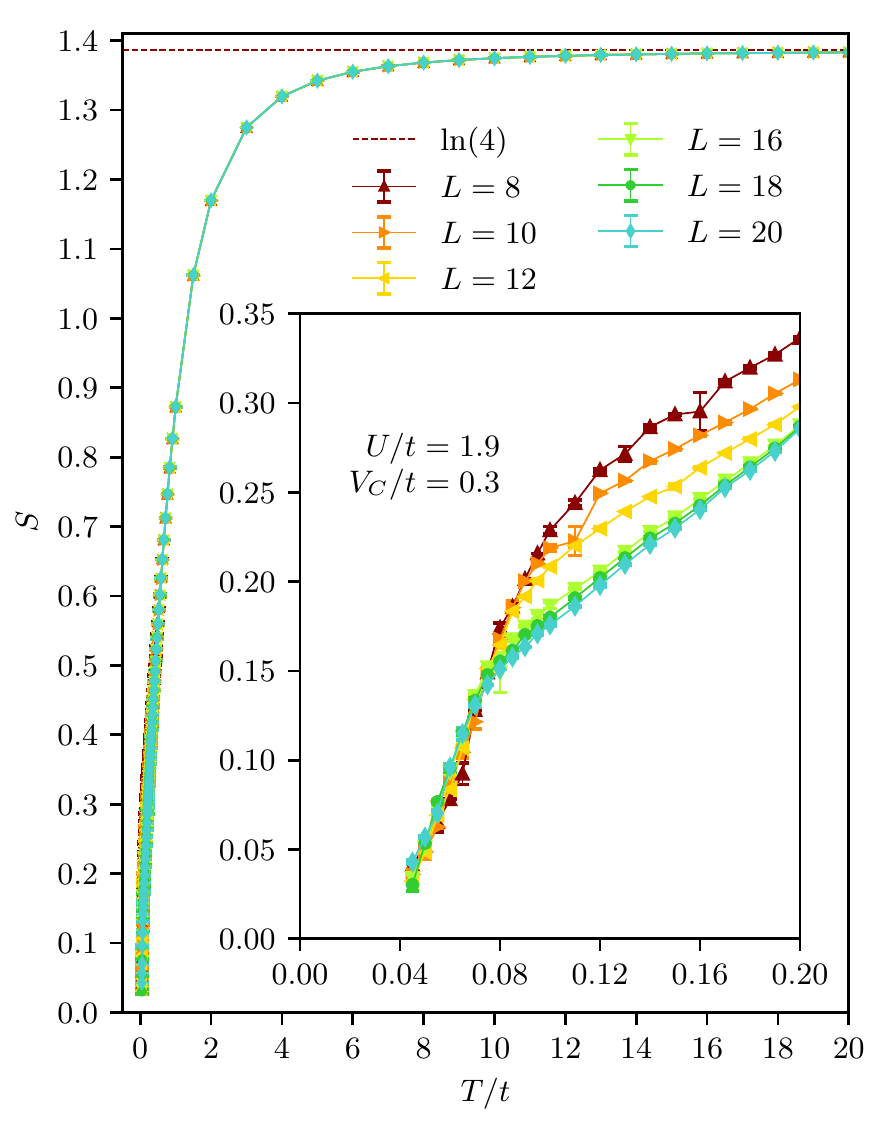}
    \caption{Temperature dependence of the entropy $S$ for the LRC-Hubbard model at $U/t=1.9$, $V_C/t=0.3$. The inset focuses on the low-temperature regime.}
    \label{Fig:Coulomb_S}
\end{figure}

For completeness, we also consider the case of LRC interactions. More specifically, we fix again $U/t=1.9$ and set $V_C/t=0.3$, corresponding to the case where the variational approach yields a discontinuity in $D$ of a similar magnitude than for the $U$-$V$ model case. In Figs.~\ref{Fig:Coulomb_D} and \ref{Fig:Coulomb_S}, we present our DQMC data for this case. See also Fig.~\ref{Fig:comparison} for a comparison to the Hubbard and U-V model.  For the LRC case, we again observe an overall enhancement of $D$ and $S$ with respect to Hubbard model. More specifically,  the presence of the additional interactions  beyond the nearest-neighbor terms leads to a weaker  enhancement of $D$ than for the $U$-$V$ model, while the low-temperature enhancement of $S$ over the Hubbard model case is larger for the LRC-Hubbard model than for the $U$-$V$ model. On the other hand, also for the LRC-Hubbard case we do not obtain any  indication for the presence or for the  onset of a discontinuity in the DQMC data for $D$, in contrast to the variational approach~\cite{Schueler19}.
Also for the LRC-Hubbard we performed additional DQMC simulations for varying values of $V_C$ within the sign-problem free regime for $U/t=1.9$. The DQMC data for the various considered quantities, $D$, $S$, $S_\mathrm{AFM}$, and $S_\mathrm{CDW}$, are shown in the right panels of Figs.~\ref{Fig:D_ext},~\ref{Fig:S_ext}, ~\ref{Fig:SAF_ext} and ~\ref{Fig:SCDW_ext}, respectively. While for the considered values of  $V_C/t$ larger than  $0.6$ we were not able to reach sufficiently down to the asymptotic low-temperature regime to see, e.g. the asymptotic low-$T$ suppression of the entropy (cf. Fig.~\ref{Fig:S_ext}), the additional DQMC data  exhibits very similar behavior and trends as for $V_C/t=0.3$ within the relevant temperature range around $T/t\approx 0.06$, throughout the accessible interaction regime. 

\section{Conclusions}\label{Sec:Conclusions}
In summary, we have examined the thermodynamic behavior of the double occupancy and the entropy of both the U-V and LRC extended Hubbard model, focusing on the low-temperature regime, for which a recent variational calculation reported the emergence of a weakly first-order MIT transition, as compared to the smooth crossover in the local Hubbard model limit~\cite{Schueler19}. In agreement with the variational calculations, we observe an overall enhancement of the double occupancy for the systems with extended interactions and we also identify an associated increase of the  entropy within the low-temperature regime.  
However, both our finite-size data as well as  carefully extrapolated TDL limit values do not provide us with any  evidence for the presence or the onset of non-continuous behavior as reported from the variational approach. Furthermore, we observe deviations from the variational calculations already for the local Hubbard model case of an order of magnitude that is significantly larger than the size of the weak discontinuities reported in Ref.~\cite{Schueler19}. This indicates that the discontinuous behavior reported previously steams from inherent limitations from the variational calculations in combination with the interpolations employed in Ref.~\cite{Schueler19}. Upon approaching the limiting values of $V=U/4$ for sign-problem free DQMC simulations for the $U$-$V$ model, we observe a (weak) peak emerging in the temperature dependence of the double occupancy. This could  be linked to the enhancement of local density fluctuations in the vicinity of the thermal critical point of the CDW order that emerges for sufficiently large values of $V$ (but beyond the DQMC-accessible interaction regime).

It would  certainly be interesting to extend these investigations on the thermodynamic effects of extended interactions in systems of correlated electrons with even larger values of the extended interaction terms, beyond the limits of sign-problem free DQMC simulations in future works, based on, e.g., tensor network or minimally entangled thermal typical state approaches. 

\section*{Acknowledgements}
We acknowledge support by the Deutsche Forschungsgemeinschaft (DFG) through Grant No. WE/3649/4-2 of the FOR 1807 and through RTG 1995, and thank the IT Center at RWTH Aachen University and the JSC Jülich for access to computing time through the JARA Center for Simulation and Data Science.

\appendix

\begin{figure}[t]
    \centering
    \includegraphics{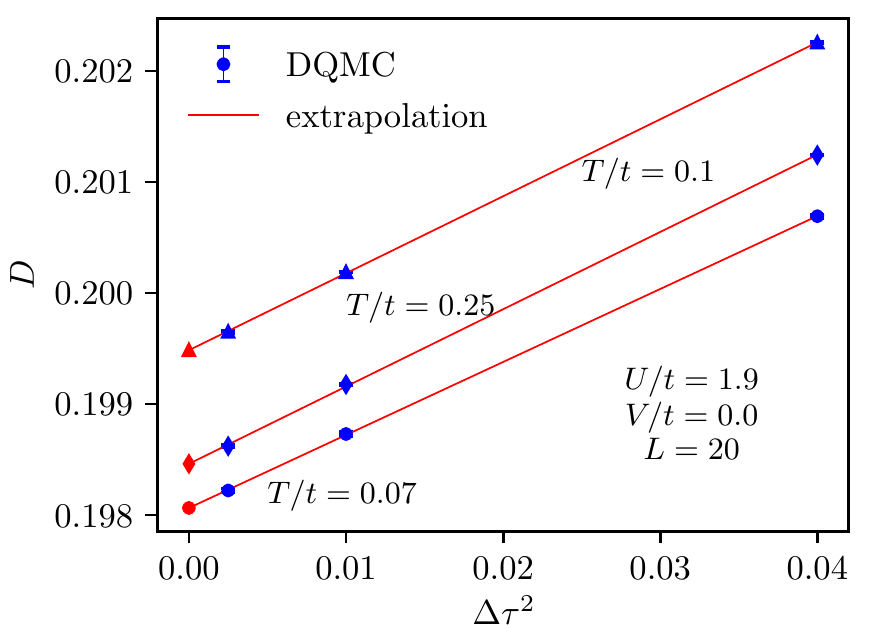}
    \caption{Trotter-discretization extrapolation of the double occupancy $D$ for the Hubbard model at $U/t=1.9$, $L=20$, for different values of the temperature
    $T$. }
    \label{Fig:dtau_D}
\end{figure}

\section{Trotter-discretization extrapolation}\label{App:A}
Based on to the hermiticity of the physical observables, the Trotter-error that arises in the discrete-time DQMC calculations due to the finite Trotter-discretization scales proportional to $\Delta\tau^2$ in the asymptotic region~\cite{Blankenbecler81}. This property 
allows us to systematically extrapolate the DQMC data to the $\Delta\tau\rightarrow 0$ limit. This process is illustrated for  representative data sets in Fig.~\ref{Fig:dtau_D} 
(access to all the obtained finite-$\Delta\tau$ DQMC data is provided via an online repository ~\cite{repo}).

\section{Finite-size extrapolation}\label{App:B}
\begin{figure}[t]
    \centering
    \includegraphics{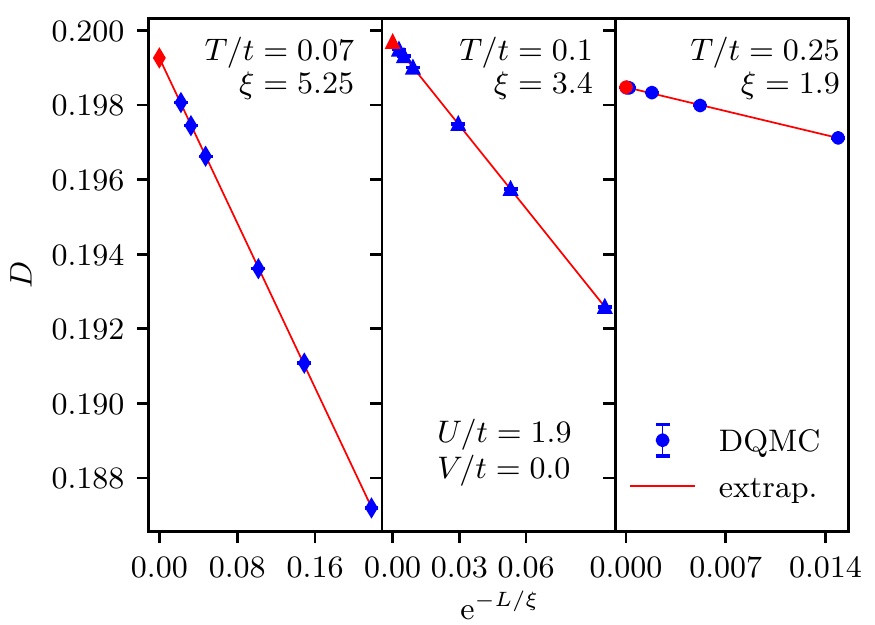}
    \caption{Finite-size extrapolation of the double occupancy $D$ for the Hubbard model at $U/t=1.9$
    for different values of the temperature $T$.}
    \label{Fig:L_D}
\end{figure}
In order to extrapolate the finite-size DQMC data to the TDL, we need to  account for the leading finite-size correction at low-temperatures in terms of a finite (correlation) length scale from  thermal fluctuations. In particular, for the double occupancy, the TDL value $D_\text{TDL}$ is obtained by fitting the finite-size data $D$ (as obtained from performing the $\Delta\tau\rightarrow 0$ extrapolation, cf. App.~\ref{App:A}) to the finite-size form $D(L)-D_\text{TDL}\propto \exp(-L/\xi)$, where $\xi$ is a $T$-dependent  parameter that quantifies the corresponding length scale. This procedure is illustrated for several representative data sets in Fig.~\ref{Fig:L_D} 
(access to all the obtained finite-$\Delta\tau$ DQMC data is provided via an online repository~\cite{repo}).

\bibliography{references.bib}

\end{document}